\documentclass[twocolumn,prl,showpacs,amsmath,amssymb]{revtex4}

\usepackage{graphicx}

\def\be{\begin{equation}}
\def\ee{\end{equation}}
\def\bea{\begin{eqnarray}}
\def\eea{\end{eqnarray}}
\def\bma{\begin{mathletters}}
\def\ema{\end{mathletters}}

\def\0{\overline{0}}

\def\q0{\underline{0}}

\def\one{\leavevmode\hbox{\small1\normalsize\kern-.33em1}}

 \def\ket#1{|#1\rangle}

\begin{document}
\title{ Security bounds for continuous variables quantum key distribution
 }

\author{Miguel Navascu\'es and Antonio Ac\'\i n}

\affiliation{ICFO-Institut de Ci\`encies Fot\`oniques, Jordi
Girona 29, Edifici Nexus II, E-08034 Barcelona, Spain
 }

\date{\today}


\begin{abstract}

Security bounds for key distribution protocols using coherent and
squeezed states and homodyne measurements are presented. These
bounds refer to (i) general attacks and (ii) collective attacks
where Eve interacts individually with the sent states, but delays
her measurement until the end of the reconciliation process. For
the case of a lossy line and coherent states, it is first proven
that a secure key distribution is possible up to 1.9 dB of losses.
For the second scenario, the security bounds are the same as for
the completely incoherent attack.

\end{abstract}


\maketitle

Quantum Cryptography, that is quantum key distribution (QKD)
followed by one-time pad, allows two honest parties to interchange
private information in a completely secure way. Quantum states
sent through an insecure channel are used to establish
correlations between the sender, Alice, and the receiver, Bob.
Since the channel is not secure, the eavesdropper, Eve, can
interact with the sent states. However, the no-cloning theorem
\cite{WZ} limits her action: she cannot produce and keep a perfect
copy of the intercepted quantum state. After this correlation
distribution, Alice and Bob employ reconciliation techniques in
order to distill from their list of classical symbols, perfectly
correlated and completely random bits about which Eve has no
information, that is a secret key. This key is later consumed for
sending
private information by means of the one-time pad. 

The first QKD protocol was introduced by Bennett and Brassard in
1984 \cite{BB84} and uses two-dimensional quantum systems, or
qubits, as information carriers. After it, other QKD protocols
were presented \cite{review}, using finite-dimensional systems as
well. More recently, it has been shown that protocols based on
continuous variables quantum systems could offer an alternative to
finite-dimensional schemes. The first of these protocols employed
squeezed states of light and homodyne measurements
\cite{GP,squeeze}. Later, a QKD protocol using coherent states and
homodyne measurements was proposed in \cite{GG} and experimentally
demonstrated in \cite{nature}. 

The security of continuous variables QKD protocols against any
type of attack has already been proven, both for the squeezed
\cite{GP} and coherent \cite{IVC} case. The bounds derived in
these works provide sufficient conditions for a secure key
distribution. There also exist restricted security proofs (see for
instance \cite{squeeze,GG,GG2}) where Eve is assumed to apply an
incoherent attack, that is (i) she interacts with the sent states
individually and in the same way and (ii) performs incoherent
measurements before the reconciliation process has started. The
corresponding bounds, then, can be seen as necessary conditions
for a secure key distribution \cite{note}. Unfortunately, there
exists a clear gap between the security conditions for general and
individual attacks, and we are far from establishing necessary and
sufficient conditions for security.

In this work, we analyze the security of QKD protocols employing
coherent or squeezed states and homodyne measurements. Using the
techniques developed in Refs. \cite{CRE,RK,DW}, we find new
security bounds for these schemes in the two following scenarios:
first, we impose no assumption on Eve's attack and derive a simple
condition for general security. Later, we assume that Eve applies
to the sent states the optimal individual interaction, but,
contrary to previous proofs, she delays her measurement until the
end of the reconciliation process. This type of attacks is
sometimes called {\sl collective}. In this second scenario, and
for the case of a lossy line, we show that the limits for key
distillation coincide with those found for incoherent attacks. To
our knowledge, this is the first situation in which it is proven
that to let Eve delay her measurement until the end of the
reconciliation process does not modifiy the security region.

By completing this work, we learn that similar results have
independently been obtained by Grosshans \cite{grosshans}.

\begin{figure}
  \includegraphics[width=8.5 cm]{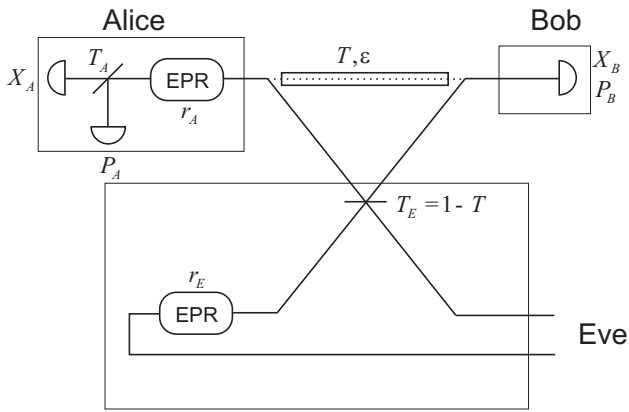}\\
  \caption{The picture shows the considered protocols.
  After Alice's effective preparation using an entangled state,
  coherent or squeezed states of light are sent to Bob, according to
  a Gaussian modulation. Eve replaces the channel by the
  entangling cloner of parameters $r_E$ and $T_E$.}\label{prot}
\end{figure}

{\sl QKD protocols:} In all the considered protocols, Alice sends
to Bob squeezed or coherent states of light modulated by a
Gaussian probability distribution. These states propagate through
a quantum channel characterized by its transmission $T$ and excess
noise $\varepsilon$. Bob randomly measures one of two quadratures,
$X$ or $P$, and communicates the chosen measurement to Alice.
Alice and Bob obtain a list of correlated real numbers from which
the key has to be extracted. There exists an entanglement-based
protocol that is completely equivalent to this prepare and measure
(P\&M) scheme \cite{GCWTG}. Indeed, Alice's preparation can be
done by measuring half of a two-mode squeezed state, of squeezing
parameter $r_A$, as shown in Fig. \ref{prot}. For instance, Alice
can measure both quadratures, $X_A$ and $P_A$, after the
beam-splitter of transmittivity $T_A=1/2$. The corresponding P\&M
scheme consists of Alice sending coherent states with a Gaussian
probability distributions of variance
$<X^2>=<P^2>=(\cosh(r)-1)/2$. On the other hand, if $T_A=1$ and
Alice chooses randomly the measured quadrature, she is effectively
sending squeezed states of squeezing parameter $\cosh(r)$ and
modulated with a Gaussian distribution of variance
$<X^2>=\sinh(r)^2/(2\cosh(r))$. This entanglement description
simplifies the theoretical analysis of the protocols, but the
obtained security bounds automatically apply to the corresponding
P\&M scheme.

It is also convenient at this point to introduce Eve's optimal
individual attack, the so-called entangling cloner. As proven in
\cite{GG2,GC}, the optimal way in which Eve can ``simulate" the
channel $(T,\varepsilon)$ is by combining into a beam-splitter of
transmittivity $T_E=1-T$, the intercepted state and half of a
two-mode squeezed state. The squeezing parameter, $r_E$, has to be
chosen such that $(1-T)\cosh r_E=1-T+\varepsilon T$.

{\sl General security proof:} Recently, powerful techniques for
the analysis of general security proofs of any QKD protocol have
been presented in \cite{CRE,RK}. In any QKD scheme, there is a
tomographic process that partly characterizes the insecure channel
connecting Alice and Bob. It allows the honest parties to evaluate
their mutual information, $I_{AB}$. Moreover, it puts a bound on
Eve's knowledge: it has been shown in \cite{CRE} that, using the
information collected during this process, one can construct a
secure reconciliation protocol that allows to extract
\begin{equation}\label{crebound}
    K= I_{AB}-\max_{\rho_{AB}\in{\cal R}}S(\rho_{AB}) ,
\end{equation}
secret bits, where ${\cal R}$ is the set of quantum states
consistent with the measured probabilities (see \cite{CRE} for
more details). Thus, this quantity represents a lower bound to the
achievable key rate, $K_{opt}\geq K$. For continuous variable
systems, if Alice and Bob monitor the channel by means of the
first and second moments of their data, the state $\rho_{AB}$ of
maximal entropy in (\ref{crebound}) has to be Gaussian
\cite{notegauss}. This is indeed the case for the attack in Fig.
\ref{prot}.

A simple calculation shows that for the same measured quadrature,
the joint probability distribution of Alice and Bob's results is
Gaussian with covariance matrix
\begin{equation}
\label{gammaab}
    \gamma_{AB}=\begin{pmatrix}T_A\cosh r_A+R_A & \sqrt{T_AT}\sinh r_A \cr
    \sqrt{T_AT}\sinh r_A & T\cosh r_A+R\cosh r_E \end{pmatrix}
    ,
\end{equation}
where $R=1-T$ and the same for $R_A$. This gives the first term in
(\ref{crebound}). The second term can be computed from $\rho_E$,
since $S(\rho_{AB})=S(\rho_E)$. Eve's state is a two-mode Gaussian
state, with covariance matrix $\gamma_E=\gamma'_E\one_2$,
\begin{equation}
\label{gammae}
    \gamma'_E=\begin{pmatrix}T\cosh r_E+R\cosh r_A & 0 \cr
    0 & \cosh r_E \end{pmatrix}
    .
\end{equation}
Using (\ref{gammaab}) and (\ref{gammae}), it is straightforward to
compute $K$ as a function of $T_A,r_A,T,\varepsilon$.

\begin{figure}
  \includegraphics[width=8 cm]{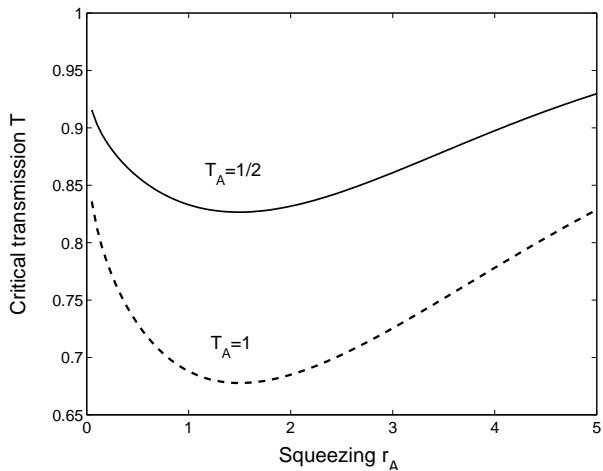}\\
  \caption{Tolerable losses as a function of the modulation
  for protocols using coherent (solid line) and squeezed
  (dashed line) states. The optimal value for both schemes
  is $r_A\approx 1.5$.}\label{gensectr}
\end{figure}

For the case of a lossy line, $\varepsilon=0$, one can numerically
see that there exists an optimal squeezing $r^{opt}_A$ for both
the coherent and squeezed case (see Fig. \ref{gensectr}). A
possible reason for this counter-intuitive result may be that $K$
is known to be a non-tight bound to the optimal key rate
\cite{CRE}. This optimal squeezing is the same for squeezed and
coherent states, $r^{opt}_A\approx 1.5$, and defines a critical
value for the tolerable losses of approximately $1.7$ and $0.83$
dB.

As discussed in \cite{CRE} it is possible to improve the bound
(\ref{crebound}) by conditioning the privacy amplification process
on a classical random variable $W$ (see \cite{CRE} for more
details), decreasing Eve's entropy. For the case of coherent
states, Alice and Bob can make public the value of the second
measured quadrature, instead of discarding it \cite{grosshans}.
This process does not modify Alice and Bob's mutual information
but changes Eve's entropy. The obtained critical transmission,
$T_c$, is now a decreasing function of the squeezing, as expected.
One can see that in the limit of high modulation,
$r_A\rightarrow\infty$,
\begin{equation}\label{secgr}
    T_c=\frac{e^2}{e^2+4} .
\end{equation}
That is, the protocol using coherent states and homodyne
measurements is secure up to 1.9 dB of losses.

{\sl Collective attack:} As said above, the bound (\ref{crebound})
is very useful because it does not make any assumption on Eve's
attack, but is known not to be tight. Moreover, it doesn't allow
to distinguish between direct and reverse reconciliation
protocols, where the (one-way) flow of information goes from Alice
to Bob and viceversa.  This distinction doesn't play any role in
QKD protocol using finite-dimensional quantum systems, but is
relevant for continuous variables protocols \cite{GG2}. Indeed,
for the case of a lossy line and incoherent attacks, the value of
the channel transmission limiting the security is equal to 1/2 for
direct reconciliation, while it goes to zero for reverse
reconciliation, for squeezed and coherent states \cite{GG2}.

In what follows, we will consider an attack where the only
assumption is that Eve applies to the intercepted states the
optimal individual interaction. In the entanglement picture, this
corresponds to the case where Alice, Bob and Eve share $N$
independent realizations of a quantum state $\ket{\Psi_{ABE}}$.
This scenario is again represented in Fig. \ref{prot}. However,
contrary to the usual incoherent attacks previously studied, Eve
delays her measurement until the end of the reconciliation
protocol. Note that this attack is clearly coherent, because Eve
can globally measure her $N$ quantum states. Moreover, she can
optimize her measurement according to all the communication
interchanged during the whole reconciliation process.

After Alice and Bob's measurements, the three parties share $N$
independent realizations of classical-classical-quantum (ccq)
correlated variables $A$, $B$ and $\ket{\psi_E}$. Under the $N$
independent realizations assumption, it has been shown in
\cite{RK,Renato} that there exists a key distillation protocol
achieving a rate
\begin{equation}\label{rkbound}
    K'=I_{AB}-\chi(A:E) ,
\end{equation}
where $\chi$ denotes the Holevo bound \cite{Holevo}. This security
condition is rather intuitive: if this quantity is positive, the
information Bob has on Alice's symbols is larger than the
classical information accessible to Eve through the quantum
channel Alice-Eve. The results of \cite{RK} prove that this
advantage can indeed be exploited for distilling a key. Moreover
they can be seen as the generalization of the results of
\cite{DW}, where Bob's information was quantum (cqq). When
considering reverse reconciliation, a similar expression holds
where $\chi(A:E)$ is replaced by $\chi(B:E)$.

For a lossy line and direct reconciliation, the critical
transmission, $T_c$, limiting the security is again a decreasing
function of the squeezing $r_A$. In the limit of very high
modulation one can see that $T_c$ is the solution to the equation
\begin{equation}\label{trcritdr}
    \frac{T_c(1-T_c)(1-T_c+T_A(2T_c-1))}{T_A+T_c-2T_AT_c}=(1-T_c)^2
    .
\end{equation}
Remarkably, $T_c=1/2$ is the searched solution $\forall\, T_A$,
i.e. we recover the same value as for the completely incoherent
attack \cite{GG}. Therefore, by increasing the modulation $r_A$,
the limiting losses value for a key distribution secure against
collective attacks tends to 3 dB.

A stronger result is obtained for reverse reconciliation. When $T$
and also $r_A$ are small, the key rate (\ref{rkbound}) goes as
\begin{equation}\label{trcritdr}
    K'\approx T_AT(\cosh r_A-1) .
\end{equation}
Therefore, there is no loss limit for reverse reconciliation
protocols. But perhaps more surprisingly, there is no need of high
modulation or squeezing for recovering the same limits as for the
completely incoherent attack! That is, a protocol using coherent
states and any modulation is secure for all line transmissions,
even if Eve is assumed to delay her measurement until the end of
the whole reconciliation process.

\begin{figure}
  \includegraphics[width=8 cm]{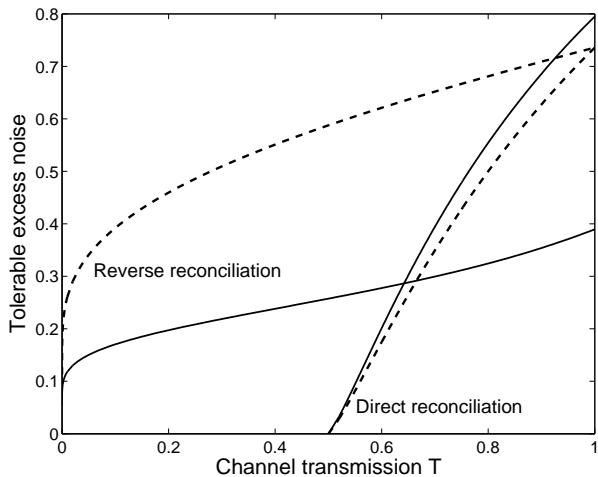}\\
  \caption{Tolerable excess noise as a function of the losses
  for reverse and direct reconciliation and squeezed (dashed line)
  and coherent (solid line) states. All the curves have been computed
  in the limit of very high modulation, $r_A\rightarrow\infty$.}
  \label{gensecen}
\end{figure}

Concerning the amount of excess noise the protocols tolerate, this
is shown in Fig. (\ref{gensecen}). For squeezed states, it is
always more convenient to employ reverse reconciliation
techniques. For the case of coherent states, direct reconciliation
turns out to be more resistant against excess noise up to a
channel transmission of $\approx 0.65$. Note that there exist
limiting values of the excess noise, $\varepsilon_c$, for which
the considered key rates are zero, independently of the modulation
and the losses. These values can be computed analytically. For
coherent states and direct reconciliation, one has that
$\varepsilon_c$ is the solution to the equation
\begin{equation}\label{encrdc}
    \frac{1}{1+\varepsilon}\left(\frac{\sqrt{1+\varepsilon}+1}
    {\sqrt{1+\varepsilon}-1}\right)^{\sqrt{1+\varepsilon}}=e^2 ,
\end{equation}
that gives $\varepsilon_c\approx 0.8$, while for reverse
reconciliation
\begin{equation}\label{encrrc}
    \varepsilon_c=\frac{1}{2}\left(\sqrt{1+\frac{16}{e^2}}-1\right)
    \approx 0.39 .
\end{equation}
In the case of squeezed states, the critical noise is equal to
$2/e\approx 0.7$ for both reconciliation protocols.

{\sl Concluding remarks:} In this work we have applied the recent
security proofs of Refs. \cite{CRE,RK,DW} to QKD protocols using
coherent and squeezed states and homodyne measurement. It has to
be clear that the obtained results provide lower bounds to the
achievable secret key rate. Thus, they represent sufficient
conditions for key distillation for the studied scenario (and
assumptions).

The first of the analyzed conditions is very powerful because does
not make any assumption on the eavesdropping attack. For a lossy
line and coherent states, it has been shown here that a secure key
distribution is possible up to 1.9 dB of losses. Existing proofs
of security work up to 1.4 \cite{IVC} and 1.6 dB \cite{GP} of
losses. Thus, our results slightly improve the known region of
general security, without requiring any squeezing.

The second type of bounds does not refer to the most general
situation, since Alice, Bob and Eve are assumed to share $N$
linearly independent copies of a quantum state $\ket{\Psi_{ABE}}$.
Consequently, we have considered the case where Eve applies the
optimal individual interaction \cite{GCWTG,GC} to any sent state.
Therefore, in this first step of her attack, Eve is assumed to
introduce no correlations among the quantum states shared by Alice
and Bob. The bounds obtained in this scenario cannot be seen as
proofs of general security. However, it is now possible to
distinguish between direct and reverse reconciliation, a relevant
issue for continuous variables quantum cryptography. Remarkably,
the obtained bounds turn out to be the same as for the fully
incoherent attack. The case of reverse reconciliation is perhaps
more surprising, since this is true for any value of Alice's
modulation. As far as we know, this is the first situation in
which it is proven that allowing Eve to delay her measurement does
not give her any significant advantage (see also
\cite{grosshans,long}).

We would like to conclude with a brief comment on the $N$
independent copies assumption required for the bound
(\ref{rkbound}). According to our results, the only possibility
left to Eve that could modify the security bounds for the fully
incoherent attack would be to introduce correlations among the
different copies of the states. Does this fact provide any
improvement on her attack? As discussed in what follows, one could
expect this not to be the case. After some channel tomography,
Alice and Bob know to share a state $\rho_{AB}^{(N)}$ such that
any single copy is a state $\rho_{AB}\in{\cal R}$, consistent with
their measured probabilities. They should assume that Eve has
tried to be as correlated as possible to their state. Since the
global state is pure, this means that Eve has optimized the
entanglement of $\ket{\Psi_{ABE}^{(N)}}$ over the splitting
$AB-E$. Thus, she has maximized the entropy of entanglement
\cite{entrent} of $\ket{\Psi_{ABE}^{(N)}}$, i.e. the entropy of
the local state $\rho_{AB}^{(N)}$, subject to the constraint that
the single-copy state is $\rho_{AB}$. This maximization naturally
leads to the $N$ independent copies assumptions, since
$S(\rho_{AB}^{(N)})\leq N\,S(\rho_{AB})=S(\rho_{AB}^{\otimes N})$.
Although far from being a proof, this simple argument, as well as
Eq. (\ref{crebound}), suggests that the best attack could consist
of Eve preparing $N$ copies of the most entropic state. If this
was true, the bounds derived from (\ref{rkbound}) would hold and
provide a necessary and sufficient condition for a secure QKD over
a lossy line using coherent (or squeezed) states and homodyne
measurements.

We acknowledge discussion with S. Iblisdir, R. Renner and
especially F. Grosshans. This work is supported by the Generalitat
de Catalunya and the Ministerio de Ciencia y Tecnolog\'\i a, under
the ``Ram\'on y Cajal" grant.


\begin{thebibliography}{99}
\bibitem{notegauss}
Since $K$ only provides a lower bound on $K_{opt}$, one cannot
conclude that the optimal attack has to be Gaussian.
\end{thebibliography}
\end{document}